\documentclass[conference]{IEEEtran}
\usepackage{cite}

\ifCLASSINFOpdf
  \usepackage[pdftex]{graphicx}
  \usepackage{subcaption}
\else
\fi
\usepackage{amsmath}
\usepackage{algorithm}
\usepackage[noend]{algpseudocode}

\usepackage{array}
\usepackage{fixltx2e}
\usepackage{url}

\usepackage[labelfont=it,textfont={it},font=small]{caption}
\usepackage{verbatim}

\title{Virtual Network Migration on the GENI Wide-Area SDN-Enabled Infrastructure}

\author{\IEEEauthorblockN{Yimeng Zhao, Samantha Lo, Ellen Zegura, Mostafa Ammar}
\IEEEauthorblockA{School of Computer Science\\
Georgia Institute of Technology\\
Email: \{yzhao389,samantha,ewz,ammar\}@cc.gatech.edu}
\and
\IEEEauthorblockN{Niky Riga}
\IEEEauthorblockA{Raytheon BBN Technologies\\
Cambridge, MA, USA\\
Email: nriga@bbn.com}}

\begin{document}
\maketitle

\begin{abstract}
A virtual network (VN) contains a collection of virtual nodes and links assigned to underlying physical resources in a network substrate. VN migration is the process of remapping a VN’s logical topology to a new set of physical resources to provide failure recovery, energy savings, or defense against attack. Providing VN migration that is transparent to running applications is a significant challenge. Efficient migration mechanisms are highly dependent on the technology deployed in the physical substrate.  Prior work has considered migration in data centers and in the PlanetLab infrastructure. However, there has been little effort targeting an SDN-enabled wide-area networking environment -- an important building block of future networking infrastructure. In this work, we are interested in the design, implementation and evaluation of VN migration in GENI as a working example of such a future network. We identify and propose techniques to address key challenges: the dynamic allocation of resources during migration,  managing hosts connected to the VN, and flow table migration sequences to minimize packet loss. We find that GENI's virtualization architecture makes transparent and efficient migration challenging. We suggest alternatives that might be adopted in GENI and are worthy of adoption by virtual network providers to facilitate migration.

\end{abstract}

\section{Introduction}


Virtualization is well-recognized as a technique to share physical resources, providing the appearance of dedicated resources and isolation from others
sharing the same physical resources. 
Virtual networks run over a physical network substrate, with an allocation of 
physical network resources (e.g., routers, switches, links, paths, or portions thereof) to the virtual network. A virtual network (VN) thus
contains a collection of virtual nodes and virtual links assigned to a subset of the underlying physical resources. A virtual link spans one 
or more physical links in the substrate, and a substrate node can host multiple virtual nodes.  


Network virtualization allows significant flexibility in network operation. Most important are the flexibility  in the VN's {\em placement} (the specific mapping of VNs elements to substrate resources \cite{VNmappingsurvey}) and VN {\em agility} (the ability to remap the VN to a different set of substrate resources over time). 
Our interest in this paper is on enabling VN agility through {\em VN migration} mechanisms. This refers to the process of remapping some or all of a VN's logical topology to a new set of physical resources. 

VN migration research considers both {\em policy}, when and why a VN is migrated, and {\em mechanism}, how a VN is migrated.
Research into VN migration policy is motivated by specific objectives. These have included: 
efficient utilization of dynamic resources \cite{fan2006dynamic,fajjari2011vnr}, recovery from 
failure \cite{tang2008efficient,gillani2012fine}, defending against attacks \cite{gillaniagile}, and reducing energy consumption \cite{MiucciVN}.

Our focus in this paper is on VN migration mechanisms. The development of such mechanisms can be quite challenging because of the  desire to make them {\it transparent} or  {\it seamless} -- informally, with minimal impact on running applications.
A further challenge is that a VN migration mechanism is highly dependent on the technology deployed in the substrate. There is, therefore, no generic mechanism that can be used universally. Previous research has developed mechanisms for VN migration in different environments.
In the data center context, Ghorbani et al. develop migration methods within their LIME architecture that 
provably meet a transparency
definition based on valid behaviors in a migration-free setting~\cite{ghorbani2014transparent}. 
In the wide-area context, Lo et al.~\cite{lo2014virtual} develop a tool for VN migration in PlanetLab~\cite{chun2003planetlab}, a
well-known shared infrastructure used for network experimentation. Their PL-VNM tool implements a migration schedule heuristic
that minimizes packet loss under ideal conditions, but in practice cannot ensure zero loss, due in part to coarse timing control
in PlanetLab. 

In this paper we focus on developing VN migration mechanisms for GENI,
a recently developed infrastructure for sharing wide-area network resources~\cite{berman2014geni}. A key technology included in GENI is software-defined networking (SDN) where the packet-processing
rules in switches are installed and modified from a logically-centralized controller~\cite{mckeown2008openflow}. SDNs offer a number of advantages
including ease of network management and the opportunity for increased innovation. 
Our focus on GENI is motivated by the fact that SDN-enabled wide-area networks are likely to become an important building block of future networking and GENI represents a fully-functional instantiation of this technology.  As such, techniques developed for GENI will have wider applicability. Further, because SDN technology is at a stage where its future can be influenced, lessons we learn about the capability of such technology in supporting network agility can have significant value on future developments.



\begin{figure}[h]
  \vskip -10pt
  \centering
  \includegraphics[width=0.5\textwidth]{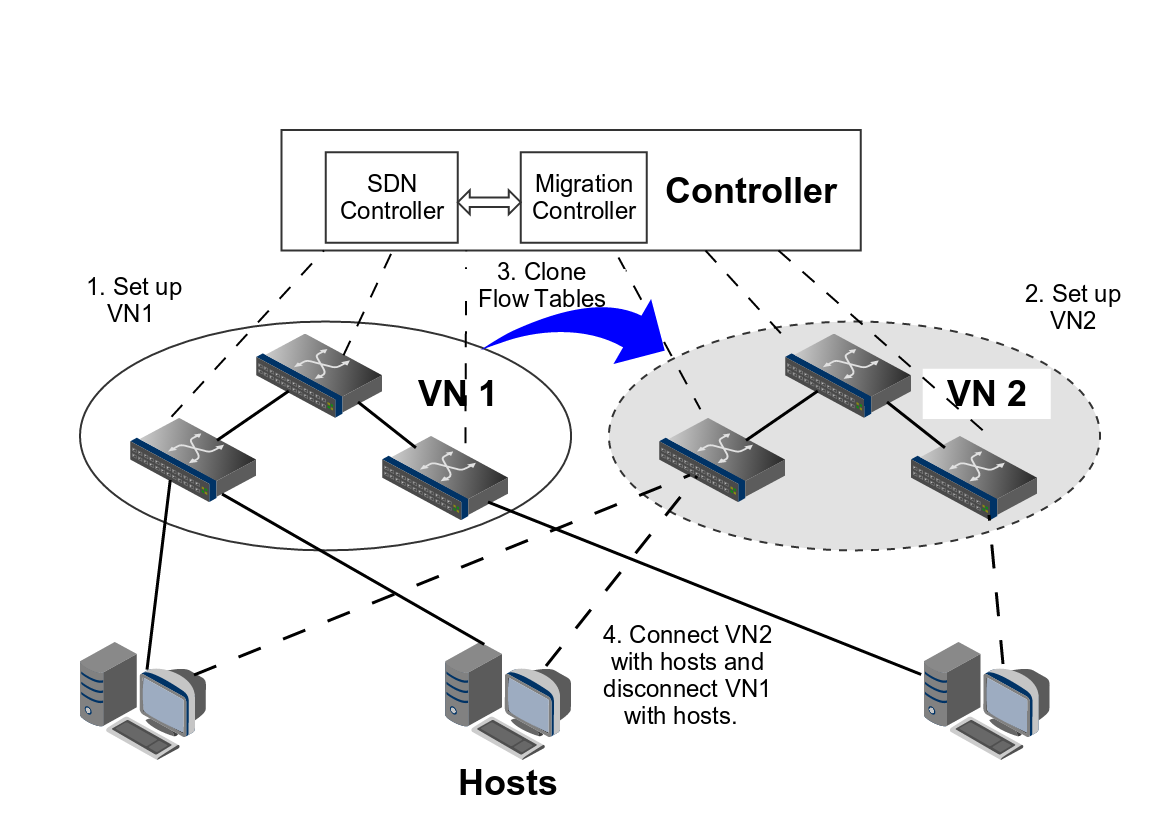}
  \vskip -10pt
  \caption{VN migration process from VN1 to VN2:In step 1, setup VN1 and connect virtual switches in VN1 to the SDN controller. In step 2, setup VN2 and connect virtual switches in VN2 to the SDN controller. In step 3, the migration controller clones flow tables from VN1 to VN2 based on the mapping. In step 4, connect VN2 with the hosts and disconnect VN1.}
  \vskip -15pt
  \label{fig:migration-steps}
\end{figure}

Our work focuses on migrating an entire VN from the initial placement to the final placement, without moving the hosts~\footnote{
Note that this model for hosts differs from the data center context where hosts are migrated with the VN. In shared
wide-area infrastructure, we assume the hosts are customer-premise equipment and remain in place when the VN moves.}. 
Figure \ref{fig:migration-steps} illustrates the migration steps assuming an SDN-enabled infrastructure. A migration controller 
interacts with the SDN controller to initialize and schedule the migration process. Prior to migration, virtual switches on the VN1 are controlled 
by the client application running on the SDN controller, and VN1 is used to deliver traffic (Step 1). When the migration starts, VN2 is setup (Step 2) and flow tables on the virtual switches in VN1 are cloned to the virtual switches in VN2 based on the mapping (Step 3). 
The migration controller issues commands to reconnect hosts from VN1 to VN2 in Step 4 and to disconnect VN1.

This paper addresses several challenges in realizing the basic VN migration steps above. In addressing these challenges we make the following contributions:
(1) We develop approaches that enable VN agility in the SDN-enabled GENI infrastructure;
(2) Develop and evaluate options for dealing with the dynamic allocation of resources inherent in migration, where the initial mapping to physical resources
is known but the future mappings necessitated by migration may not be; 
(3) Propose an approach for managing the hosts that connect to the VN and will remain in place when the VN migrates; 
(4) Develop techniques to mitigate the disruption caused by live VN migration, to minimize packet loss observed by the application in the data plane and to maintain the topological view in the control plane (as observed by the application running on the SDN controller).
We carefully manage process steps and flow table migration sequences to achieve this;
(5) Evaluate, using an implementation running on GENI, how the performance of live VN migrations as a function of design decisions and network parameters; and 
(6) Expose some limitations of the GENI infrastructure and propose approaches to their mitigation.

The remainder of the paper is structured as follows. In Section \ref{sec:background} we develop a framework for enabling VN agility within the context of the GENI substrate technology. We highlight a decision regarding VN migration related to the allocation of resources in or across
GENI slices. Section \ref{sec:migration-challenges} proposes mechanisms to address the challenges associated with VN migration on GENI to meet the goals of efficiently and transparency and our solutions. We develop a controller architecture and describe the deployment of our VN migration mechanism within GENI in Section \ref{sec:implementation}.
In Section \ref{sec:performance-evaluation} we present results from experiments conducted using our prototype with the aim of evaluating its performance. Section \ref{sec:suggestion} 
discusses GENI limitations that were exposed by our research and approaches to address them. Related Work is covered in Section \ref{sec:related-work}. We conclude the paper in Section VIII.

\section{Enabling Full VN Agility in GENI}\label{sec:background}

GENI is relatively mature shared infrastructure. As such GENI provides support for
for sharing, isolation, resource allocation,  and multiple physical substrate ``owners''.
GENI, however, was not designed specifically to support our desired transparent and efficient VN agility. Our work, therefore, involves the development and evaluation of options to support VN agility within GENI. This section explores some critical aspects of providing agility support.
Because GENI uses its own unique terminology, 
Table \ref{tb:geni-concepts} summarizes how the general VN terminology maps to GENI terms.

\begin{table}[!t]
\renewcommand{\arraystretch}{1.3}
\begin{footnotesize}
\caption{GENI Context vs. Virtual Components}
\label{tb:geni-concepts}
\centering
\footnotesize
\begin{tabular}{|c|c|}
\hline
Component & GENI Context\\ 
\hline\hline
Substrate networks & GENI testbed\\
\hline
Virtual Network(s) & GENI slice\\
\hline
Physical location & GENI aggregate\\
\hline
Virtual links within a VN & LANs\\
\hline
Virtual links between VNs & Shared VLAN\\
\hline
Virtual links connecting different physical locations & Stitched links\\
\hline
Mapping between VN to physical substrate & Rspec file\\
\hline
\end{tabular}
\end{footnotesize}
\vskip -10pt
\end{table}

\subsection{Allocating VNs to Slice(s)}\label{vn-to-slice-section}
\begin{figure}
  \centering
  \includegraphics[width=0.5\textwidth]{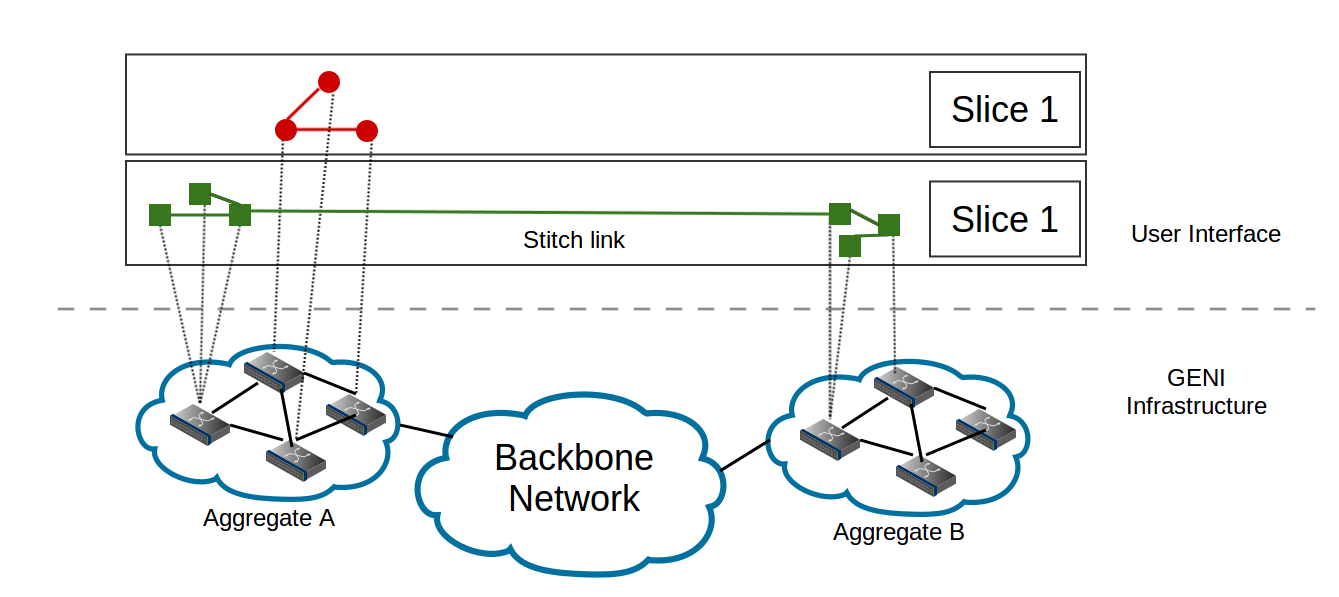}
  \vskip -10pt
  \caption{An example of GENI architecture}
  \vskip -15pt
  \label{fig:geni-arch}
\end{figure}

GENI is a ``sliced'' platform that allows concurrent experiments on shared infrastructure. A GENI slice is a unit that contains all the resources for an experiment, 
including computing resources and network links. This is already a form of network virtualization used primarily to isolate experiments in GENI. In a real-world GENI-like substrate, slicing would be used to isolate commercial network providers sharing the same physical substrate. 
 Slices exist in one or more aggregates; each aggregate is an independent collection of physical 
resources often operated by the same entity (e.g., a specific university). 
Figure \ref{fig:geni-arch} shows two slices on the GENI infrastructure. Slice 1 has three virtual nodes in Aggregate A, while Slice 2 has six virtual nodes across Aggregate A and Aggregate B connected with stitched link through the backbone network. Each slice is an isolated environment where virtual nodes and virtual links can be added. Each virtual node in a slice can be accessed by the user who creates the slice with corresponding SSH key.

Slices are meant to be deployed for the long term and are thus not agile. To enable agility, VNs will need to be deployed within slices as an additional layer of virtualization.
We consider two options for mapping VNs to slices with an eye to migration. The first option is to build all VNs (original and future) and hosts for migration within the same slice. This approach
follows the common usage model for GENI to include all resources for an experiment within a single slice. However, this option has three disadvantages: 
1) There is no clear isolation between the different VNs. 2) Most GENI resources cannot be modified after the reservation. Once resources are reserved on a slice, 
no partial modification (e.g, add a virtual link or a virtual node) is allowed. In the case of migration, this restriction requires us to reserve all resources for hosts and 
VNs, including those that will be migrated to in the future, at the outset. 3) When a VN or a host fails, we need to rebuild the whole virtual topology.

Alternatively, it is possible to allocate a single VN to a slice, starting with the original VN and later allocating a VN to migrate to. Deploying a VN on one slice is straightforward. 
The challenge for deploying and migrating a VN between two slices is caused by the difficulty to enable the inter-slice communication during migration. We cannot create a virtual link to connect virtual components in different slices directly. Instead, we can set up a VLAN, 
a broadcast domain at the data link layer, to connect virtual components in different slices. All virtual components in the same VLAN will receive the broadcasting packets even when the virtual components are in separate slices. Compared with deploying all VNs on one slice, this second design provides clear separation between VNs and gives more flexibility in resource reservation. We can reserve one VN first and create another VN when needed. However, it complicates the virtual topology during migration. We will talk further about shared VLANs and migration in Section \ref{sec:migration-challenges}.  

\subsection{Mapping Virtual Switches to Physical Machines}
GENI uses a Resource Specification (RSpec) document to describe a requested virtual topology and its physical substrate, including ID of the virtual machines (VM),  virtual network interface configuration, VLAN tags assigned to links, and corresponding hardware information. The RSpec is submitted with the GENI aggregate manager API to aggregates in order to reserve resources. The requested virtual topology is translated to a request Rspec file by Rspec generation tools, and GENI aggregates automatically allocate resources to the requested VN based on the Rspec file.

While GENI aggregate's automatic assignment of resources can meet the requirements of most experiments, it may be necessary to have the flexibility of mapping virtual nodes to specific physical resources for VN migration research. Although Rspec generating tools do not directly support resource assignment, we are able to map a virtual node to a specific machine by manually modifying the request Rspec. The Omni tool \cite{omni} provides commands to obtain information about all reserved and available resources at a specific aggregate, including machine ID, machine status, hardware types, and OS types. We can locate the component name for a specific physical machine in the resource information returned by Omni and copy its component ID to the request Rspec file to achieve a fixed mapping.

\subsection{Assigning VNs to Substrates}
In VN migration, it might be necessary to migrate between different physical substrates, or aggregates in GENI terminology. A GENI aggregate comprises a set of resources under common control including storage resources, computing nodes, and OpenFlow switches \cite{berman2014geni}. Experimenters can reserve multiple aggregates within the same slice and connect them with stitched links (See Figure \ref{fig:geni-arch}). It is also possible to allocate each VN to a different aggregate and connect them with both shared VLAN and stitched links. We will show how to use shared VLAN and stitched links together in Section \ref{sec:migration-challenges}.

\section{Dealing with VN Migration Challenges}\label{sec:migration-challenges}
After VN agility is enabled in GENI as discussed in the previous section, we are still faced with several challenges as we strive to meet the goals of efficiency and transparency. 
In this section we investigate three distinct challenges and propose mechanisms to deal with them. The challenges are: (1) how to manage inter-slice communication that connects the hosts to both VNs temporarily during the migration; (2) how to minimize 
packet loss by scheduling the flow table migration sequence; and (3) how to provide a seamless interface to SDN applications during
and after migration. The first challenge is specific to a sliced platform like GENI and the other two challenges can be generalized to other SDN environment. In the section following this one we use the solutions for each challenge to inform the design of a migration controller architecture. 

\subsection{Inter-slice Connection}\label{inter-slice-connection-section}
We described two VN-to-Slice allocation options in Section \ref{vn-to-slice-section}. In the first option, all hosts, the old VN, and the new VN are located within the same slice. We will not discuss the first design in detail since it follows the common usage of the GENI testbed. We will focus on the second design, where the old VN, the new VN, and the hosts are assigned to three different slices. The challenge in the second design is to direct traffic from one slice to another given the current GENI constraints which do not support virtual links between slices. It should be noted that the dynamic tunnel implemented in LIME \cite{ghorbani2014transparent} does not apply for our case. The tunnel uses the control plane for data links and cannot guarantee performance such as bandwidth and latency. Moreover, the control plane is a shared channel on GENI and should not be used to send a large amount of data.

\subsubsection{Broadcasting problem in a virtualization environment}
\begin{figure}[h]
  \vskip -10pt
  \centering
  \includegraphics[width=0.4\textwidth]{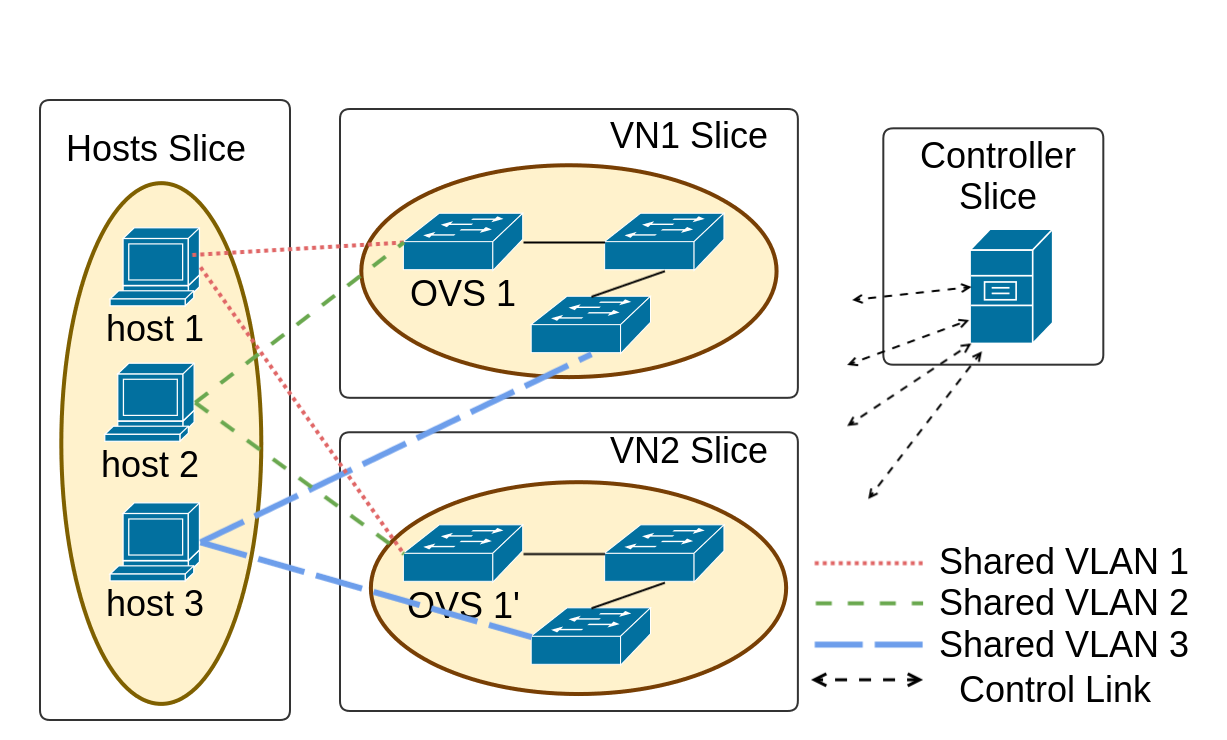}
  \vskip -5pt
  \caption{An example of using shared VLAN to connect the host and the VNs}
  \vskip -10pt
  \label{fig:wo-gw}
\end{figure}

To enable inter-slice communication, it may seem natural to use a shared VLAN to connect a host slice to VN slices. The traffic is broadcast within the same VLAN, no matter in which slice a switch is located. The connection/disconnection of the VNs with the hosts are controlled by turning up/down the network interfaces on virtual switches. Figure \ref{fig:wo-gw} presents an example of this approach. The topology includes three hosts, the old VN (VN1), the new VN (VN2), and the controller slice. In our virtual topology, each host connects to both VN1 and VN2 with a shared VLAN. When host1 sends data to host2, the data will be broadcast to both OVS1 and OVS1'. When VN1 is in use, the network interfaces of OVS1 is up and the network interfaces of OVS1' is down. After the migration, we redirect traffic from VN1 to VN2 by turning down the interfaces of OVS1 and turning up the interfaces of OVS1'. 

Unfortunately, this approach can violate the correctness of the migration in a virtualized environment. GENI uses XEN \cite{barham2003xen} as a virtual machine monitor to allow multiple virtual machines to share the same hardware resources. Xen only allows a privileged virtual machine called domain 0 to access the physical Network Interface Card (NIC). Domain 0 communicates with other virtual machines through a set of back-end interfaces. All the packets destined to a virtual machine will be first transferred to domain 0 and then destined to the virtual machine. The packets stored in domain 0 are not dropped when the network interfaces in the virtual machine is turned down. When the virtual network interface goes up again, these buffered packets will be copied from domain 0 memory to the receiver virtual machine's memory.

\begin{figure}[h]
  \vskip -10pt
  \centering
  \includegraphics[width=0.4\textwidth]{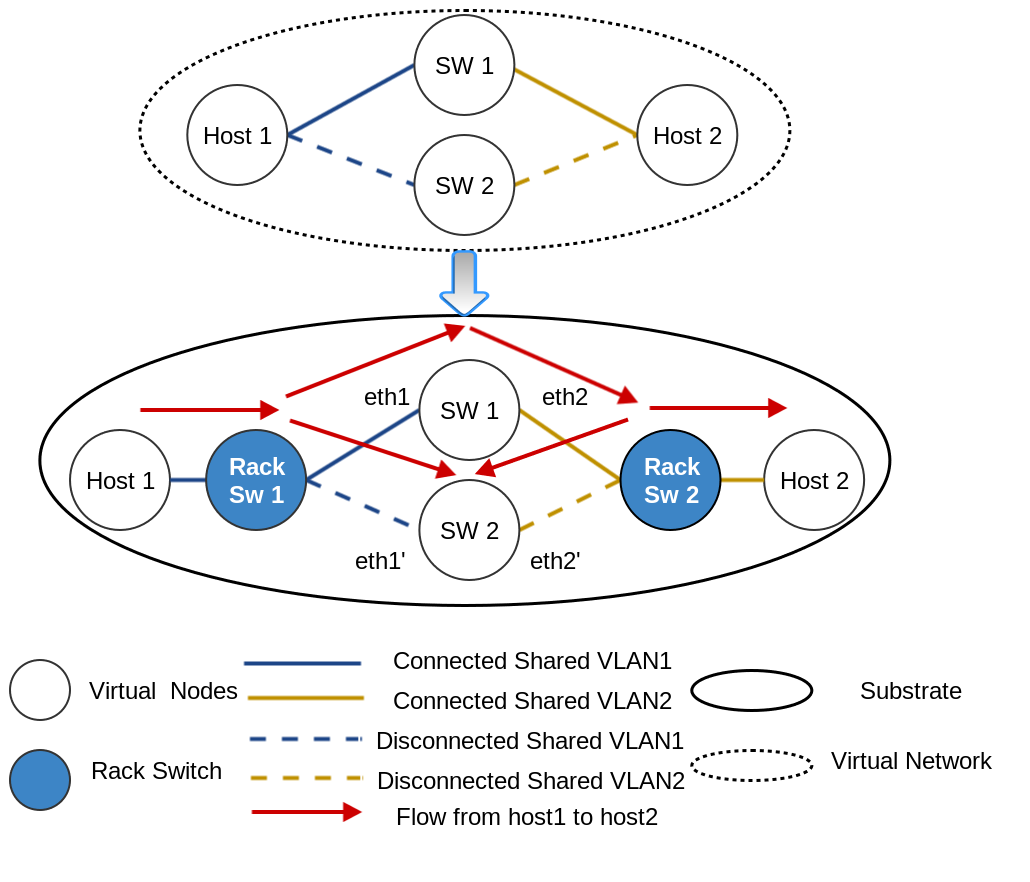}
  \vskip -10pt
  \caption{an example of a VN and its substrate}
  \vskip -10pt
  \label{fig:broadcast-problem}
\end{figure}

We illustrate why this small number of buffered packets can be a problem through a one-node VN example as shown in Figure \ref{fig:broadcast-problem}. In our virtual topology, host1 connects two switches with shared VLAN1 and host2 connects two switches with shared VLAN2. In actual substrate network, there is a rack switch residing in the shared VLAN to broadcast packets to all switches in the same VLAN. Before migration, we connect VN1 with the hosts by turning up the network interfaces eth1 and eth2 and disconnect VN2 by turning down the network interfaces eth1' and eth2'. The data from host1 to host2 is broadcast by Rack{\_}SW1 to eth1 and eth2, and then broadcast by Rack{\_}SW2 to eth2' and host2. Although we turn down the virtual network interface eth1' and eth2', a small number of packets are still stored in the XEN domain 0. 

During the migration, we switch from VN1 to VN2 by turning up eth1' and eth2' and turning down eth1 and eth2. Previously buffered packets in domain 0 are transferred through eth1' and eth2' to the virtual machine that hosts SW2. These packets have the same matching fields (e.g, same source and destination IP) but request different actions(e.g, send through different ports). In the standard SDN implementation of the learning switch, this is considered as an error. The switch will install rules to drop all packets for several seconds, resulting in a much longer disconnection time than normal migration process. In the worst case, when conflicting rules are installed on the openflow switch, the switch may stop forwarding packets, which requires manual configuration to recover.
 
\subsubsection{Mitigate the broadcasting problem -- gateway design}
\begin{figure}
  \vskip -10pt
  \centering
  \includegraphics[width=0.4\textwidth]{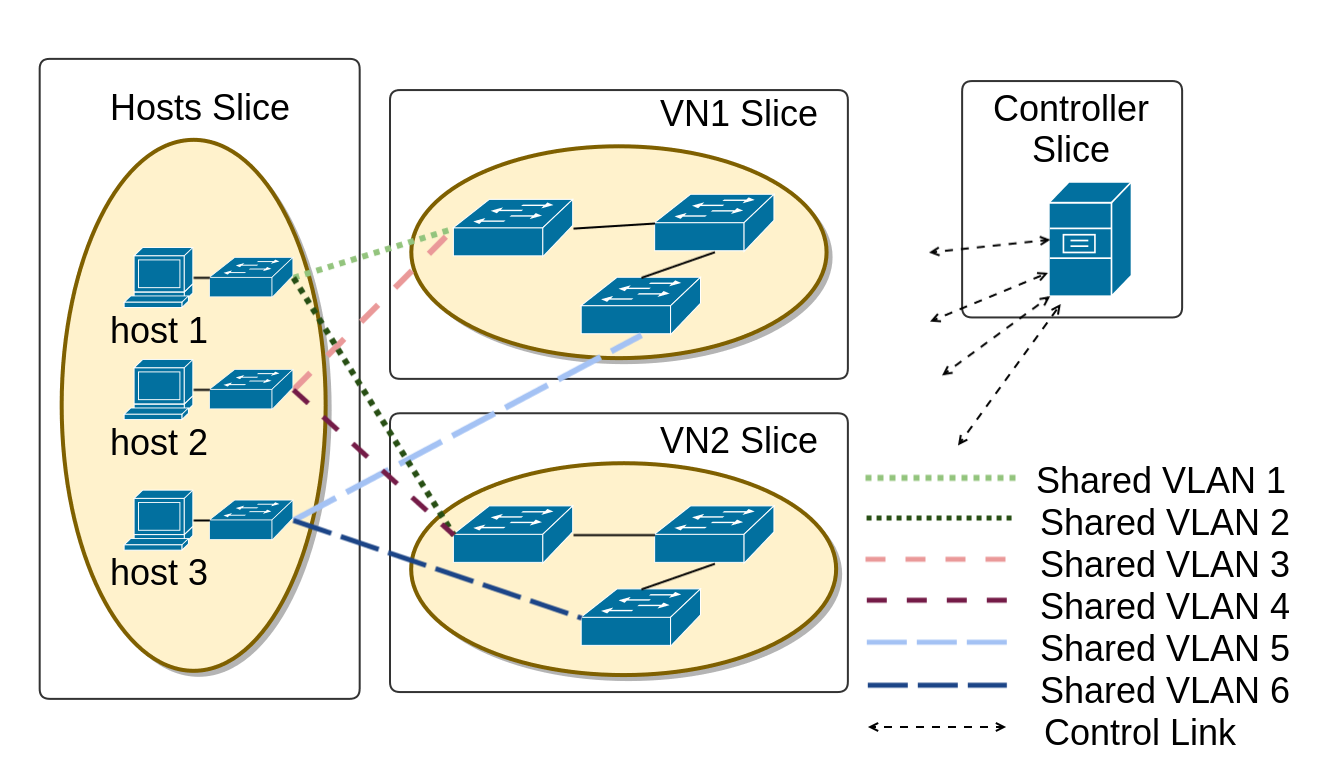}
  \vskip -10pt
  \caption{Gateway design}
  \label{fig:gw-topo}
\end{figure}

To avoid the broadcasting problem,  we propose a gateway design which establishes additional SDN switches as `gateways' to switch traffic from the old VN to the new VN. The gateways are layer 2 devices that sit between hosts and VNs, hiding changes in VNs from end hosts. Figure \ref{fig:gw-topo} presents an example of the gateway design that enables migration within the same aggregate. Each host is connected with a gateway, and each gateway uses two different shared VLANs to connect to the two VNs. The gateway switch is responsible for forwarding packets from hosts to a certain VN. In the process of VN migration, the migration controller issues commands to the gateway switches, asking them to redirect traffic from VN1 to VN2 after all flow tables in VN1 are cloned to VN2. The controller sends SDN commands to the gateway switches to update the flow tables, redirecting traffic from VN1 to VN2.

\begin{figure}
  \centering
  \includegraphics[width=0.45\textwidth, height=55mm]{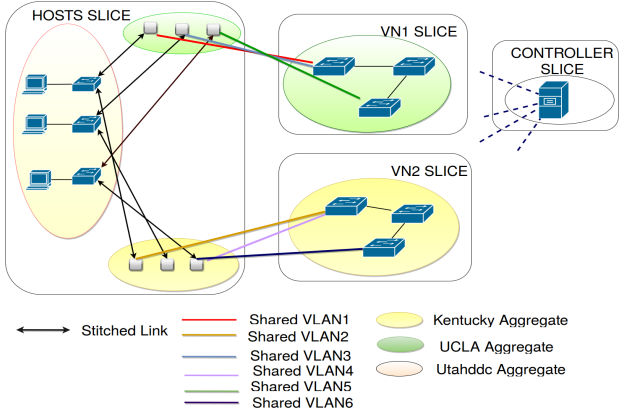}
  \caption{Topology to enable cross-aggregate migration}
  \vskip -20pt
  \label{fig:gw-topo-stitch}
\end{figure}

The gateway design can be extended to enable migration across substrates. As mentioned earlier, we use GENI aggregates to represent substrates in different locations. Virtual components in different aggregates are connected with a special link called stitched links. Unfortunately, a stitched link cannot be part of a shared VLAN. We use additional nodes to serve as a bridge to connect stitched links and shared VLANs. A cross-aggregates example is shown in Figure \ref{fig:gw-topo-stitch}. The hosts, VN1, and VN2 are located in three different aggregates. GENI does not provide inter-slice stitched links to connect gateway switches in the host slice with SDN switches in two VNs directly. To connect gateways with VN1, we put three more additional nodes in the host slice. These three nodes are in the same aggregate with VN1 and we use the stitched link to connect them with the gateway switches. Then we use shared VLANs to connect those three nodes to the virtual switches in VN1. Those three additional nodes serve as a bridge to connect the host slice with VN1. We do the same to connect the hosts with VN2.

\subsection{Minimizing Packet Loss}\label{scheduling-section}
In our migration mechanism, packet loss may occur when the migration controller issues commands to gateway switches to disconnect the old VN and reconnect the new VN. In a traditional network without SDN features, unicast Reverse Path Forwarding (uRPF) \cite{dalal1978reverse} in strict mode drops traffic received on an interface that is not used to forward the return traffic. We illustrate why VN migration always introduces packets loss in symmetric routing through a two-node topology.
 
\begin{figure}
  \centering
  \includegraphics[width=0.9\linewidth]{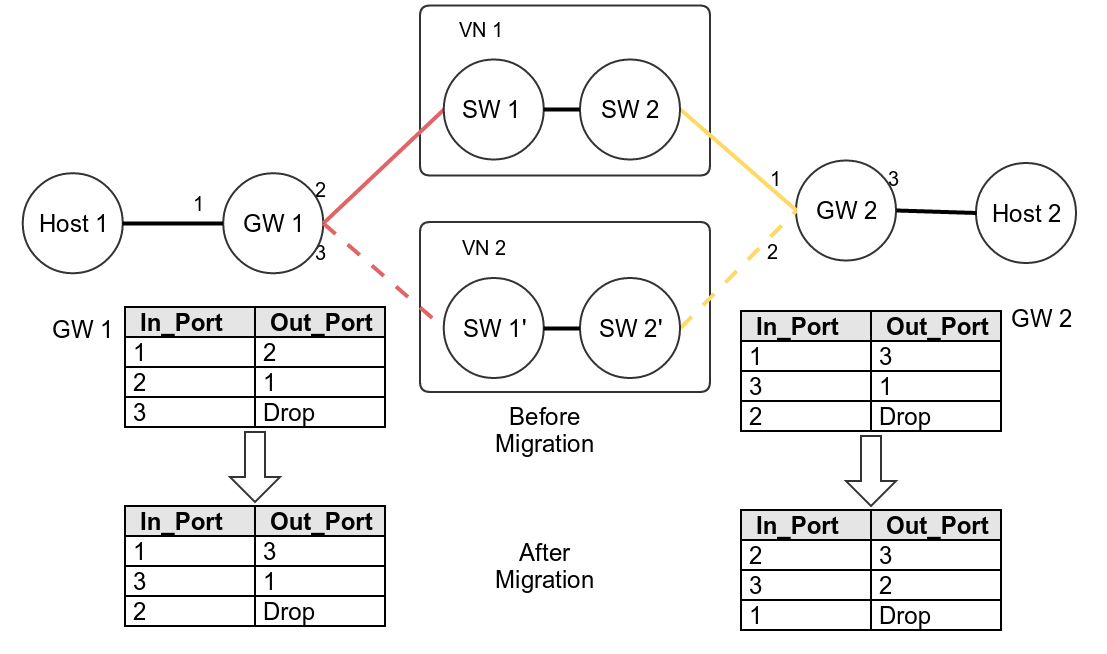}
  \caption{The topology of two-node VN on GENI}
  \vskip -20pt
  \label{fig:scheduling-problem}
\end{figure}

In Figure \ref{fig:scheduling-problem}, there are two hosts and two VNs, each VN containing two virtual nodes. Each host connects to both VN1 and VN2 through a gateway switch. We define f\textsubscript{1,2} as the traffic flow from host1 to host2, and f\textsubscript{2,1} as the traffic flow from host2 to host1. We migrate the virtual network from VN1 to VN2. Before migration, GW1 directs f\textsubscript{1,2} from in-port 1 to out-port 2, directs f\textsubscript{2,1} from in-port 2 to in-port 1, and drops any traffic from in-port 3 to disconnect VN2. The same applies for GW2 to control traffic from/to host2. When the migration begins, our migration controller issues commands to GW1 and GW2 and updates their flow tables to redirect traffic from VN1 to VN2. We assume GW1 finishes update at time t\textsubscript{1,2} and GW2 finishes at time t\textsubscript{2,1}. We define d\textsubscript{1} as the latency from GW1 to GW2 and d\textsubscript{2} as the latency from GW2 to GW1. The data rate of f\textsubscript{1,2} is r\textsubscript{1} and the data rate of f\textsubscript{2,1} is r\textsubscript{2}. We therefore calculate the number of dropped data c\textsubscript{1,2} for f\textsubscript{1,2} and c\textsubscript{2,1} for f\textsubscript{2,1} as follows: 
\[
  c_{1,2}=\begin{cases}
               (t_{2,1}-t_{1,2})-d_1)\times r_1, &\text{if } t_{2,1}-t_{1,2}\geq d_1\ \\
               (t_{1,2}-t_{2,1})+d_1)\times r_1, &\text{otherwise} 
            \end{cases}
\]

\[
  c_{2,1}=\begin{cases}
               (t_{1,2}-t_{2,1})-d_2)\times r_2, &\text{if } t_{1,2}-t_{2,1}\geq d_2\ \\
               (t_{2,1}-t_{1,2})+d_2)\times r_2, &\text{otherwise} 
            \end{cases}
\]
It is obvious that $t_{2,1}-t_{1,2} = d_1(d_1 >= 0)$ and $t_{1,2}-t_{2,1} = d_2(d_2 >=0)$ cannot be both satisfied. At least one of c\textsubscript{1,2} and c\textsubscript{2,1} is larger than 0, which means additional packet loss is unavoidable in this setting.
 
\subsubsection{Flow Migration Sequence}
SDN shows promise to enable lossless migration with an optimized sequence of rule installation. We propose a scheduling sequence to remove the additional packet drop introduced by VN migration. Algorithm \ref{alg:traffic-redirection} shows pseudocode for the traffic redirection process. We install rules to let traffic coming from the new VN to go through the gateway switches. Then we update rules on gateway switches to direct traffic from hosts to the new VN. Finally, we insert drop rules to disconnect the old VN. By following this sequence, we avoid dropping packets buffered in the old VN.

\begin{algorithm}
\caption{Traffic Redirection Algorithm}\label{alg:traffic-redirection}
\begin{algorithmic}[1]
\For{$gateway \in gatewayList$}
  \State $Ports_h \gets $Ports on gateway that point to hosts
  \For{$Port \in Ports_h$}
  \State install new rule $r$ where $r.inPort$ = $PortToVN2$ and $r.outPort$ = $Port$
  \EndFor
\EndFor
\For{$gateway \in gatewayList$}
  \For{$Port \in Ports_h$}
  \State update rule $r$ set $r.outPort$ = $PortToVN2$ where $r.outPort$ = $PortToVN1$ and $r.inPort$ = $Port$ 
  \EndFor
\EndFor
\For{$gateway \in gatewayList$}
  \For{$Port \in Ports_h$}
  \State update rule $r$ set $r.action$ = $dropPkt$ where $r.outPort$ = $PortToVN1$ and $r.inPort$ = $Port$ 
  \EndFor
\EndFor

\end{algorithmic}
\end{algorithm}

Algorithm \ref{alg:traffic-redirection} also applies to partial VN migration when only part of the old VN is remapped to different physical machines. In the partial VN migration, the traffic redirection occurs at the neighboring nodes of the partial network instead of the gateways. In this case, all neighboring nodes of the partial network should be treated as gateway switches and the same algorithm can be applied to minimize the packet loss.

\subsubsection{Remote Scheduling Methods}
We have two implementations for issuing migration commands to disconnect the old VN and connect the new VN. The first option is to control VN connection by turning up/down network interfaces using SSH sessions. We refer this type of scheduling as SSH scheduling. With this scheduling method, there is a lag between the time when the command is issued and the time when the command is actually executed in the remote node. Besides, GENI requires SSH key-based authentication before executing commands on a node, which might lead to a longer lag time.

The second implementation, called OpenFlow-message scheduling, redirects traffic by installing flows on gateways based on the OpenFlow messages from the migration controller. This method does not support complicated operations such as executing a monitoring script. We expect this method to be faster than SSH scheduling because it does not introduce authentication overhead.

\subsection{Seamless Migration}
Our migration mechanism should ensure the illusion of a consistent VN for the client SDN applications during and after the migration. We discuss possible inconsistencies in the migration process and our solutions.

\subsubsection{Topology Changes}
As mentioned in Section \ref{sec:migration-challenges}, both the old and the new VN are connected during the migration. To present the client SDN applications with a consistent view of a single VN, our migration controller intercepts all OpenFlow events, changes the datapath ID of the events based on the mapping between the old and the new VN, and passes the modified events to the client SDN applications. No events about the topology changes in the VN2 should be passed to the client SDN applications. 
     
\subsubsection{Switch State Changes}
A switch maintains information about its ports and flow tables. Ideally, a switch should present the same switch Information including datapath ID, serial number and ports in the old and the new VN. Unfortunately, GENI does not allow users to assign virtual network interfaces to the virtual switch and the ports number are randomly assigned during reservation stage. It is highly likely that the virtual switch has different ports status in the new VN. Since the flow tables contain the port information, our migration controller modifies the flow tables based on the port mapping when it clones flow tables from old switches to new switches. 

\section{Migration Controller Architecture}\label{sec:implementation}

\begin{figure}
  \vskip -5pt
  \centering
  \includegraphics[width=0.5\textwidth]{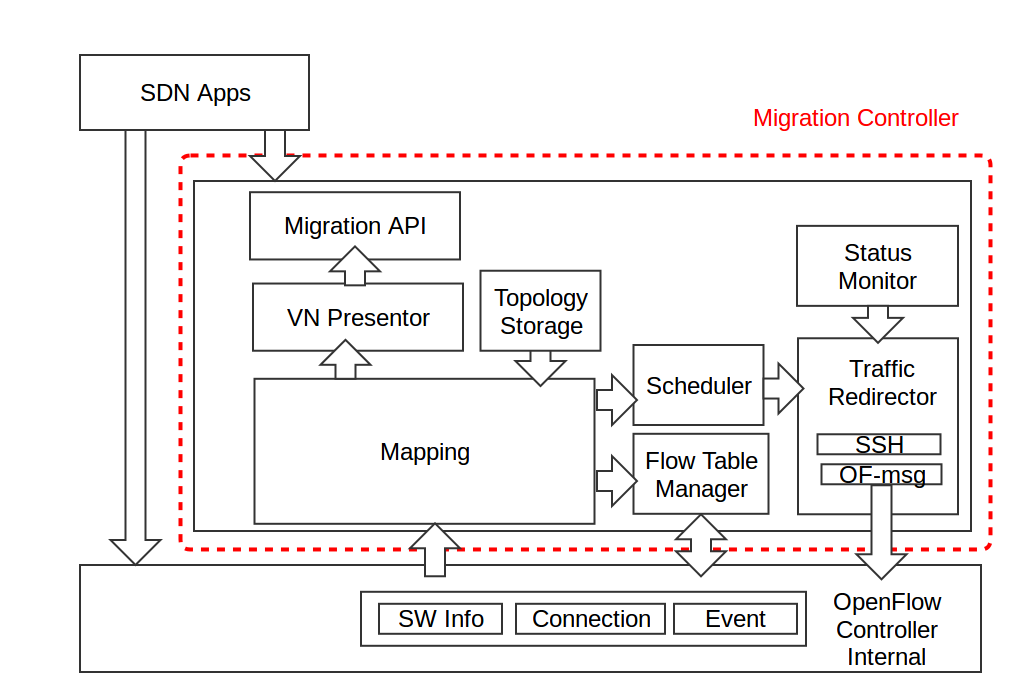}
  \vskip -5pt
  \caption{Migration controller architecture}
  \vskip -10pt
  \label{fig:advanced-controller}
\end{figure}

The migration controller stands in the center of our migration architecture and is responsible for the migration process. It clones the flow tables from the old switches to the new switches, schedules the migration sequences and switches traffic between VNs. We implement our migration controller on GENI using the POX controller platform \cite{pox-controller}. The migration controller runs on the POX controller while other client applications keep operating normally. The controller architecture is shown in Figure \ref{fig:advanced-controller}.  

\textbf{Mapping Module}: specifies how to map the switches in the old VN to the switches in the new VN. It also includes mapping of the virtual network interfaces in the old switches and in the new switches. When reserving resources on GENI, we cannot specify virtual network interfaces in the request Rspec file and GENI aggregate arbitrarily assigns virtual network interfaces to VMs. We need to query the virtual network interface corresponding to a certain IP address and store that information in the Mapping Module.

\textbf{Flow Table Manager}: When a request for VN migration is initiated, the Flow Table Manager polls switches that are affected by migration, translates flow tables from the old VN based on the mapping information stored in the Mapping Module, and installs the flows into the new switches. 

\textbf{Scheduler}: calculates the sequence of rule installation based on our traffic redirection algorithm to minimize the packet loss. 

\textbf{Traffic Redirector}: After all flows are successfully installed, the Flow Table Manager notifies the Traffic Redirector to generate traffic redirection commands. The traffic Redirector retrieves the sequence of rule installation from the Scheduler and redirects the traffic from the old VN to the new VN.


\textbf{VN Presenter}: intercepts events from switches, translates them based on mapping information from the Mapping Module, and presents a consistent VN topology to client applications. This module hides all migration process from clients.

\textbf{Status Monitor}: collects dynamic network statistics and decides where and when to migrate based on the VN placement algorithm. Our focus is to migration mechanisms, thus we have not implemented the Status Monitor.

\textbf{Migration API}: Migration APIs are similar to OpenFlow controller APIs so that client applications adapt to the new APIs easily. The migration APIs allow client SDN applications to configure migration parameters such as migration destinations and triggering requirements. The client SDN applications should use migration API to retrieve virtual switch information, the connections to virtual switches, and events from virtual switches to get a consistent view of the VN.

\section{Performance Evaluation}\label{sec:performance-evaluation}
In this section, we evaluate the performance of our migration mechanism in terms of the migration time, packet loss during migration, latency, and the controller overhead. More evaluation results can be found in the accompanying technical report \cite{}.

\subsection{Migration Time}
\subsubsection{SSH vs. OpenFlow-messaging}
\begin{figure}
  \vskip -10pt
  \centering
  \includegraphics[width=0.49\textwidth, height=50mm]{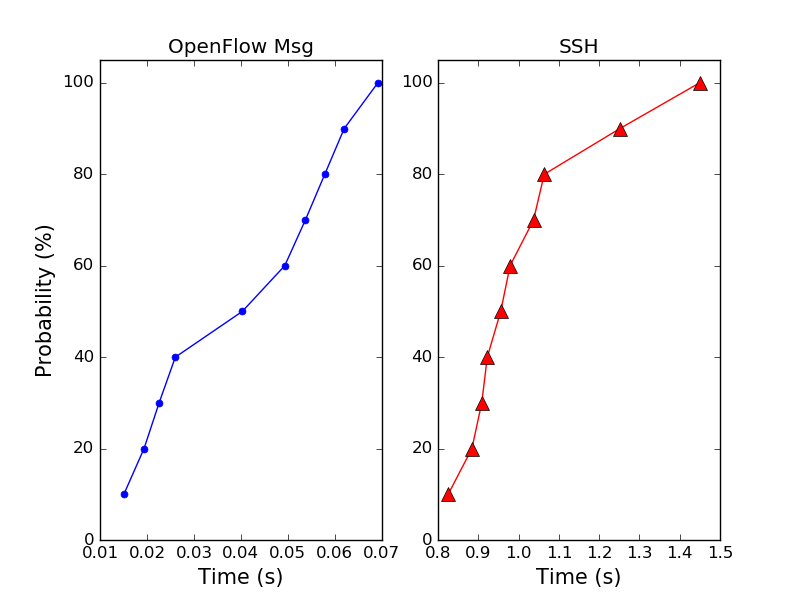}
  \vskip -5pt
  \caption{CDF of time difference between command issued and execution}
  \vskip -15pt
  \label{fig:of-vs-ssh}
\end{figure}

We evaluate whether OpenFlow-message scheduling performs better than SSH scheduling in terms of the time difference between command issue and command execution. We use our migration controller to issue command to turn down/up a network interface using SSH session for 50 times. We repeat the same experiments with OpenFlow-message scheduling. Figure \ref{fig:of-vs-ssh} shows the CDF of the time difference between the time when migration commands are issued by the controller and the time when the migration finishes with SSH scheduling and OpenFlow-message scheduling. All OpenFlow-message scheduling completes within 0.1 second, but about 50\% of SSH scheduling takes 1s or longer to finish. This confirms our earlier assessment that OpenFlow-Message scheduling is much faster and has lower variance than SSH scheduling.
 
\subsubsection{Migration Duration}
\begin{figure}
  \vskip -10pt
  \centering
  \includegraphics[width=0.4\textwidth, height=50mm]{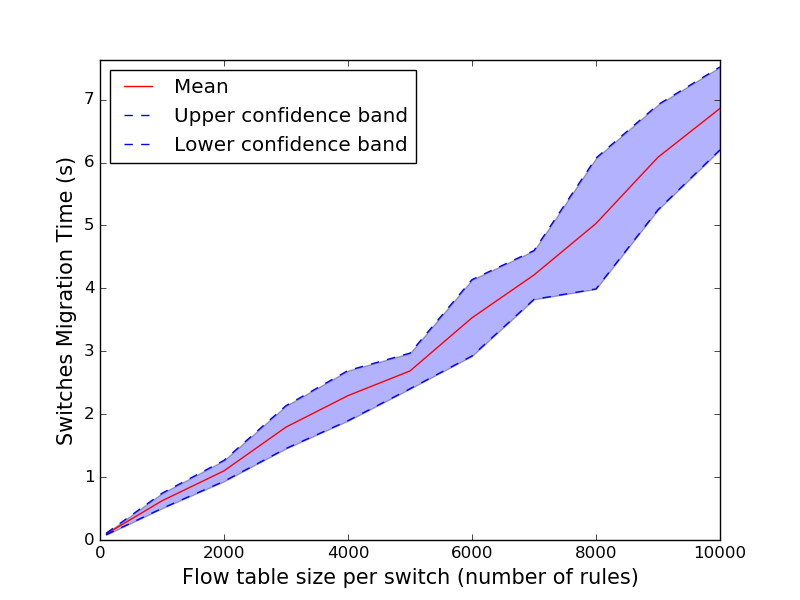}
  \vskip -5pt
  \caption{Migration time as flow table size per switch grows}
  \vskip -15pt
  \label{fig:mt-flowsize}
\end{figure}

Figure \ref{fig:mt-flowsize} shows how migration time changes as the flow table size grows with 95\% confidence level upper and lower bands. The migration time is negligible when flow table size is small. It takes less than 1s to finish the migration when the flow table size is smaller than 1000. The migration duration increases roughly linearly with the number of rules per switch and can take 7s when there are 10,000 rules. The migration time depends on the number of rules and the number of switches but is independent of the topology.    

\subsection{Packet Loss During Migration}
\subsubsection{Move a Complete VN within a Substrate}
We build a prototype of the basic migration controller using the POX controller platform \cite{pox-controller} and evaluate its performance through experiments on the topology illustrated in Figure \ref{fig:gw-topo} with three hosts and six virtual switches. All virtual switches are Open vSwitches \cite{pfaff2009extending}. We use iperf to generate UDP traffic for 10 seconds between all pairs of hosts and migrate VN from its initial position to final position at time t=5s. We vary the data sending rate to see whether our migration controller works well in relatively high data rate. We perform three sets of experiments: (a) a baseline experiment where no migration occurs, (b) migration with symmetric routing, where traffic redirection commands are issued at the same time by controller, and (c) migration with asymmetric routing, where traffic redirection commands are issued in an optimized sequence. We repeat the experiments for 30 times for each data rate and measure the migration time and data loss rate. 

\begin{figure}
  \centering
  \begin{subfigure}{0.45\textwidth}
    \centering
    \includegraphics[width=0.8\textwidth]{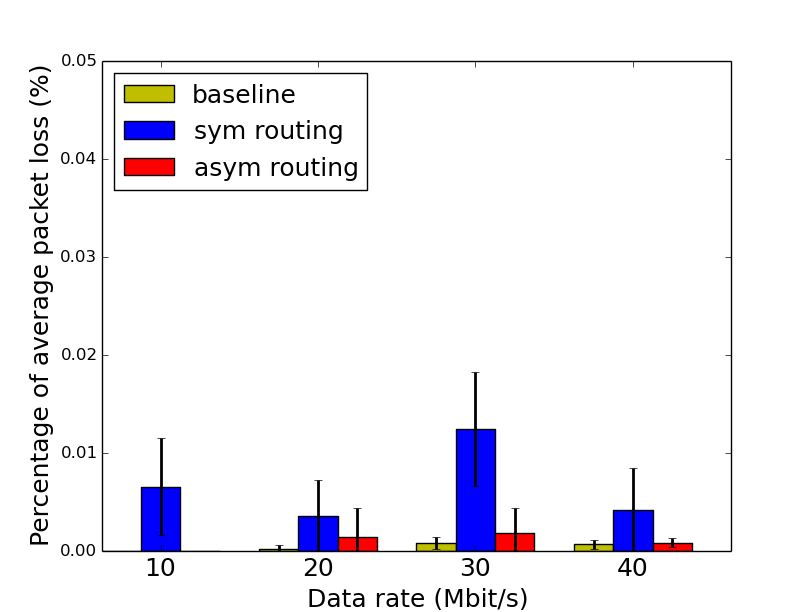}
    \caption{From host1 to host3}
    \label{fig:h1-h3-drop}
  \end{subfigure}

  \begin{subfigure}{0.45\textwidth}
    \centering
    \includegraphics[width=0.8\textwidth]{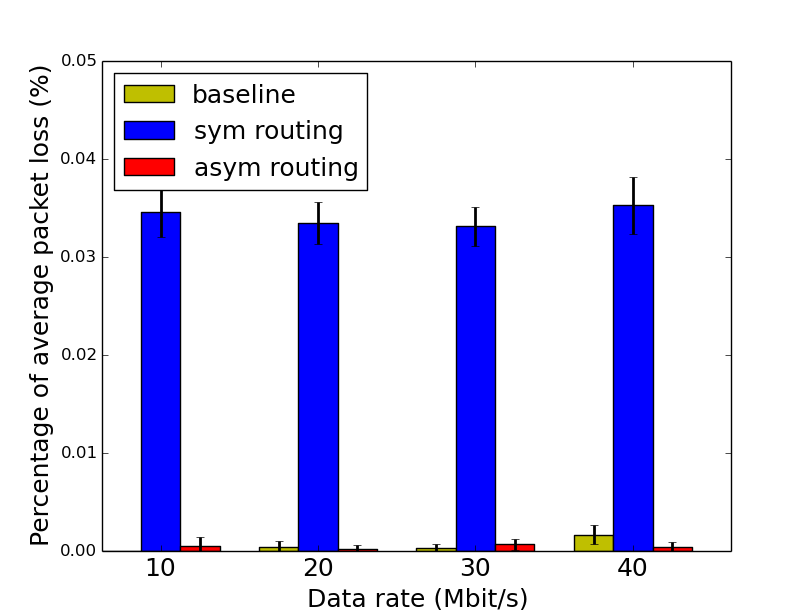}
    \caption{From host3 to host1}
    \vskip -5pt
    \label{fig:h3-h1-drop}
   \end{subfigure}

   \caption{Average packet loss as percentage of total packets (baseline, symmetric routing, optimized scheduling) at different data rates for the flow between host1 and host3}
   \vskip -15pt
   \label{fig:avg-loss}
\end{figure}

We only present results for forwarding and reverse flows between host1 and host3 due to space constraints. In Figure \ref{fig:avg-loss}, the percentage of average packet loss on y-axis is based on the measurement of UDP traffic for 10s, and the x-axis shows baseline experiment, symmetric routing, and asymmetric routing for different data sending rates. For both the forwarding and the reserve flows, the packet loss rate in asymmetric routing is almost the same with that in a migration-free setting. It demonstrate that asymmetric routing prevents hosts from experiencing significant increase in packet loss during migration. 

\begin{figure}
  \vskip -5pt
  \centering
  \includegraphics[width=0.4\textwidth, height=50mm]{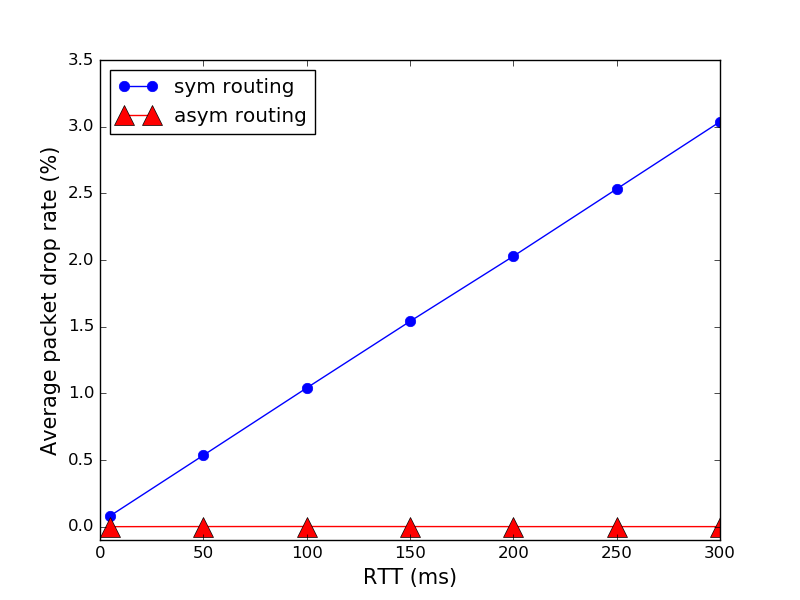}
  \caption{The performance of symmetric routing and asymmetric routing for different RTTs}
  \vskip -15pt
  \label{fig:loss-rtt}
\end{figure}

\subsubsection{Impact of RTT}
During the migration, packet loss occurs when packets buffered in the old VN are dropped by gateway switches because of the traffic redirection. As shown in Figure \ref{fig:loss-rtt}, the performance of the symmetric routing is much worse than that of the asymmetric routing, especially when the flow table size is large. The average packet drop rate for symmetric routing increases linearly with the increase of the RTT while the packet drop rate for asymmetric routing is very close to zero for any RTT values.  

\begin{figure}
  \centering
  \includegraphics[width=0.4\textwidth, height=50mm]{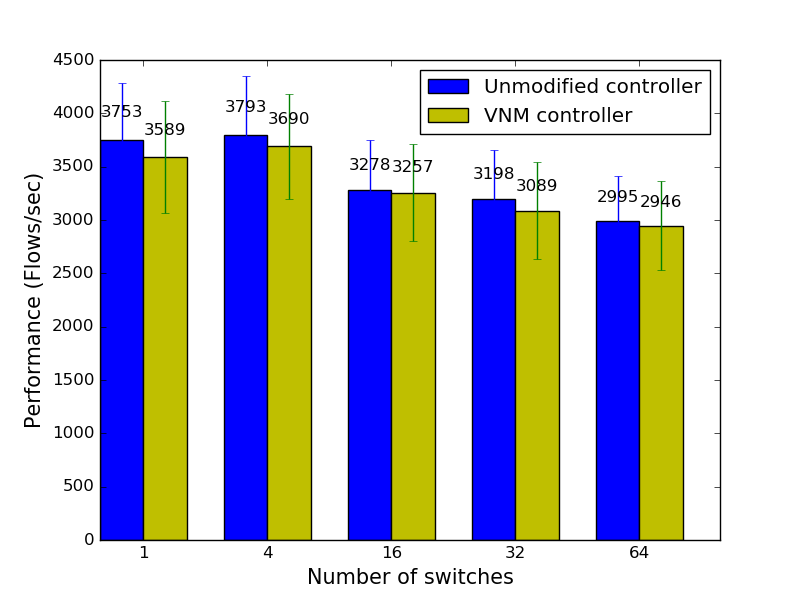}
  \caption{Controller performance for different switch numbers}
  \vskip -15pt
  \label{fig:ctrl-throughput}
\end{figure}

\subsection{Control-Plane Overhead}
Our migration controller intercepts the events, modifies the datapath ID based on the mapping between the old and the new virtual switches, and passes the new events to the client application. These operations cause overhead at the controller.
 
To evaluate the controller performance, we use the cbench program \cite{sherwood2010cbench}, which creates and sends a large number of OpenFlow messages to the controller. Figure \ref{fig:ctrl-throughput} shows performance of the unmodified POX controller and our migration controller from one switch to 64 switches. The y-axis shows the number of flows that a switch can handle within a second. Our migration controller processes roughly 3\% fewer flows per second than the unmodified controller does.

\section{Mitigating GENI Limitations}\label{sec:suggestion}
When we started our work, our goal was to design, implement, and evaluate an efficient VN migration mechanism in GENI as an example of a future SDN-enabled wide-area network.  While it is possible to deploy virtual networks on GENI and use proper remote scheduling implementation to enable live migration, we observe that some GENI limitations complicate the design. These constraints are not only particular to our VN migration research, but may also apply to other types of experimentation. We summarize the features that are not well supported by GENI. This will aid in future GENI development and also in informing the designs of GENI-inspired SDN-enabled wide area infrastructure.

\subsection{Interaction with the Substrate Network}
GENI deploys virtualization architectures to share physical resources for simultaneous experiments. It provides multiple models of virtualization to cater for different levels of performance and isolation requirements. However, experimenters are only free to select from the provided virtualization models, and do not have the privilege to modify the model or build an alternative one. In particular, we have the following constraints if our experimentation explores the interaction between the virtualized architectures and the substrate networks.

\paragraph{Little knowledge about substrate networks}
Under the current GENI context, we only have access to limited information about the substrate network such as the geographical information about GENI aggregates and VM load on each physical machine. Without sufficient real-time information about the physical substrate, it is difficult or impossible to implement an algorithm that has interaction with the substrate network. For example, some VN migration research may require real-time statistics about the substrate network to determine when to trigger the migration and where to migrate a VN. More generally, to support experimentation where the placement of the virtual topology is affected by the performance of the substrate network, we expect GENI to expose more network statistics such as link utilization, throughput, and latency.

\paragraph{Difficulty in debugging}
The virtualization techniques are deployed on GENI to support simultaneous experiments on limited physical infrastructure. However, virtualized architecture not only implies a trade-off between performance and isolation, but also makes debugging challenging. The virtualization architecture may bring unexpected problems, and the limited access to physical substrate further increases the difficulty in debugging. In our VN migration research, we had a hard time finding the cause of the duplicated packets when shared VLANs are used. We can only debug by observing the traffic in virtual topology and infer what is happening in the physical substrate. We expect GENI to develop efficient debugging tools to make the debugging process easier. Besides, it is impossible to debug without a deep understanding of the mechanisms (e.g., how shared VLAN works). Most GENI tutorials only introduce how to use their features. It would be helpful if GENI can include more architecture design of GENI features in its tutorials. 

\paragraph{No control of substrate networks} 
We have flexibility to assign bandwidth to our virtual links in the reservation stage, but we cannot adjust parameters for the substrate network. Therefore, it is difficult to evaluate an algorithm with bandwidth constraints. This constraint makes it difficult to observe how dynamics in physical substrate such as changes in bandwidth or latency can affect the performance of a virtualized architecture. 

\subsection{Multi-domain Network Research}
In GENI, a slice is a unit that contains all computing and networking resources for an experiment. The common usage of GENI is to reserve all resources for an experiment within an slice and isolate different slices. One possible design for multi-domain network experiment is to place all domains within the same slice and build different administrative controllers to handle different domains. The disadvantage is obvious: there is no isolation among domains, and we are unable to add more domains to dynamically scale up the networks. Alternatively, we can place one domain on one slice with isolated administration. To enable inter-slice communication, we need to use shared VLANs to connect slices, which complicate the virtual topology and makes it difficult to scale up. Neither of the two designs are ideal solutions for experiments that involves multiple domains.   

\subsection{Dynamic Resource Reservation}
The GENI platform requires experimenters to reserve all resources on GENI slices before running their experiments. Most GENI resource does not provide flexibility to partially modify the resources. In our work, we take advantage of the shared VLAN feature to make resource reservation more dynamic. This resource reservation method requires the experimenters to consider which virtual links in the first slice should be converted to shared VLAN at the beginning when they design their experiments. Each design is particular to a specific topology: whenever we need a new virtual topology, we need to reconsider the shared VLAN. The restriction in resource reservation makes it difficult to scale up an experiment.  

\section{Related Work}\label{sec:related-work}
 Some of the VN embedding solutions suggest reconfiguration or remapping of the VN\cite{fan2006dynamic,fajjari2011vnr,tang2008efficient,gillani2012fine,gillaniagile}. However, all of those works use simulation to demonstrate the effectiveness of their solutions. It remains a challenging task for network researchers to move their experiments to a real infrastructure when there is a lack of effective migration mechanism. 

There has been some work addressing the challenges of VN migration in a real infrastructure. Prior work \cite{lo2014virtual} proposes an orchestration mechanism for VN migration on PlanetLab, using the same technology to move a single virtual router without disrupting current traffic traversing the virtual network presented in \cite{wang2008virtual}. Other work \cite{pisa2010openflow,ghorbani2014transparent} shows how to migrate VN within software defined networks. Pisa et al. considers the basic migration scenario for migrating virtual network in traditional network and software defined network\cite{pisa2010openflow}. Ghorbani et al. \cite{ghorbani2014transparent} move the whole SDN network with the hosts in the data center context. It concentrates on low level configuration including packet-in events, traffic statistics and rule timeout to handle correctness violation \cite{ghorbani2014towards}. 

\section{Conclusion}\label{sec:conclusion}
In this paper we consider the design, implementation and evaluation of virtual network migration on GENI as a working example of a future SDN-enabled wide-area infrastructure. VN migration adds network agility to the repertoire of network functions and provides a mechanism that enables the deployment of important policies for resource management, energy conservation and attack defense.  We show how agility can be enabled on top of GENI's slicing approach, enumerate and address challenges in the design of efficient VN migration mechanisms, and develop and deploy an implementation of a controller architecture that achieves our VN migration objectives. We perform a set of experiments that help us understand the implications of various design decisions and network parameters on VN migration performance. Our work also exposes some limitations on the current design of GENI.  




\scriptsize
\bibliographystyle{IEEEtran}
\bibliography{IEEEabrv,paper}

\end{document}